\documentclass[12pt]{amsart}
\usepackage{amssymb}
\usepackage{amsmath,amsthm,amsfonts}
\usepackage[applemac]{inputenc}
\usepackage{graphicx}
\usepackage{bbm}
\usepackage{pdfsync}
\usepackage{fancyhdr}
\usepackage{mathrsfs}
\usepackage{slashed}
\usepackage{color}
\usepackage{xcolor}
\usepackage[breaklinks,colorlinks,backref]{hyperref}

\hypersetup{
colorlinks, %set true if you want colored links
linktoc=all, %set to all if you want both sections and subsections linked
linkcolor=blue, %choose some color if you want links to stand out
citecolor=hyptxt,
urlcolor=blue}
\hypersetup{
citebordercolor=,
filebordercolor=green,
linkbordercolor=blue
}
\definecolor{hyptxt}{rgb}{0.7, 0.4, 0.9}

\usepackage{bibtopic}

\usepackage[full]{textcomp} % to get the right copyright, etc.
\usepackage{fbb} % osf for text, lining for math
\usepackage[varqu,varl]{inconsolata}% typewriter
\usepackage[libertine,bigdelims,vvarbb]{newtxmath}
\usepackage[cal=boondoxo]{mathalfa}% less slanted than STIX
\usepackage[T1]{fontenc}
\useosf
\newtheorem{prop}{Proposition}[section]

\newcommand{\beprop}{\begin{prop}}
\newcommand{\enprop}{\end{prop}}
\newcommand{\bprf}{\begin{proof}}
\newcommand{\eprf}{\end{proof}}

\newcommand{\ket}[1]{|\kern.3ex#1\kern.3ex\rangle}
\newcommand{\bra}[1]{\langle\kern.3ex #1 \kern.3ex|}
\newcommand{\scalar}[2]{\langle\kern.3ex #1 \kern.3ex|\kern.3ex#2\kern.3ex\rangle}

\def\ud{\mathrm{d}}

\def\ud{\mathrm{d}}
\definecolor{hervecolor}{rgb}{0.8,0,0.7}

\numberwithin{equation}{section}

\def\1{\mbox{I\hspace{-.15em}1}}

\def\b{\begin{equation}}
\def\e{\end{equation}}

\begin{document}
%\date{\today}
\title{Krein space quantization and \\ New Quantum  Algorithms}
\author{ M. V. Takook
%\\ Shahriar Rouhani $^{2}$
}
\address{ \emph{ APC, UMR 7164, Universit\'e Paris Cit\'e, 75205 Paris, France; } \newline
%\emph{ Elevated Quantum Abacus Company }
}
\email{ takook@apc.in2p3.fr}
%; shahriarrouhani@gmail.com}
%\footnote{\emph{ APC, UMR 7164}, \emph{Universit\'e de Paris}, \emph{75205 Paris, France}, takook@apc.in2p3.fr}
\maketitle
%\vspace{15pt}
\begin{abstract}
Krein space quantization and the ambient space formalism have been successfully applied to address challenges in quantum geometry (e.g., quantum gravity) and the axiomatic formulation of quantum Yang-Mills theory, including phenomena such as color confinement and the mass gap. Building on these advancements, we aim to extend these methods to develop novel quantum algorithms for quantum computation, particularly targeting underdetermined or ill-conditioned linear systems of equations, as well as quantum systems characterized by non-unitary evolution and open quantum dynamics. This approach represents a significant step beyond commonly used techniques, such as Quantum Singular Value Decomposition, Sz.-Nagy dilation, and Unitary Operator Decomposition. The proposed algorithm has the potential to establish a unified framework for quantum algorithms.
\end{abstract}
%\vspace{0.5cm}
%{\it Proposed PACS numbers}: 04.62.+v, 98.80.Cq, 12.10.Dm
%\vspace{0.5cm}

%\tableofcontents

%%%%%%%%%%%%%%%%%% %%%%%%%%%%%%%%%%%%%%%%%%%%%%%%%%%%%%%

\section{Introduction}

Theoretical physics aids in the development of technology, while advancements in technology, in turn, enhance our understanding and refinement of theoretical models and physical concepts. For instance, quantum computing has enabled us to reinterpret the fundamental concepts of particles and waves as quantum states, known as qubits, and to recognize these states as physical entities through which computations are performed \cite{taja}. In this work, we aim to focus on developing and analyzing new quantum algorithms that are conceptually and mathematically inspired by ideas from quantum Yang-Mills theory and quantum gravity, as recently presented \cite{taqg0,ta231,taqg,taga22}. By briefly recalling the problems that arise in gauge theory and quantum geometry and reviewing the proposed approaches to their solutions, we compare them with the challenges in existing quantum computing algorithms. Through this examination of the similarities between these two areas, we aim to foster a deeper understanding of both.

Krein space quantization \cite{gareta00} and ambient space formalism \cite{ta14} have been utilized to analyze axiomatic quantum Yang-Mills theory \cite{ta231,taga22} and quantum geometry, leading to the definition of a complete quantum space of states, the quantum state of the universe, and its unitary evolution \cite{taqg0,taqg}. In these cases, the Hilbert space is not a complete space and must be replaced with Krein space. The ambient space formalism enables the definition of an observer-independent perspective, which is crucial in quantum geometry. The observer effect remains one of the most challenging issues in quantum measurement and quantum technology, particularly in quantum computing.

It is sometimes stated that the evolution of quantum geometry or the gravitational field is non-unitary. Similarly, the minimally coupled scalar field in de Sitter space encounters this issue \cite{gareta00}. It is well known that this scalar field can be interpreted as the conformal component of the metric tensor field \cite{anmo,anmamo}, or the geometry of spacetime, thereby leading to the problem of non-unitary evolution in quantum geometric fields \cite{taqg0}. From general relativity, we know that the source of geometry or gravity is the matter and radiation fields \cite{swgr}. Within the framework of quantum field theory in curved spacetime \cite{bida}, we can apply the weak gravitational approximation and treat the quantum field as an approximately closed quantum system, thereby enabling the construction of a complete Hilbert space; for the specific case of de Sitter curved spacetime, see \cite{tagahu}.

The weak field approximation method is not valid when applied to quantum geometry. In the weak matter-radiation field approximation, the curvature of spacetime or gravitational field would vanish, which contradicts the inherent nature of quantum geometry. Thus, in quantum geometry, the matter-radiation field cannot be neglected. Quantum fields are embedded within the geometry and contribute to its structure, making quantum geometry an open system. Consequently, it becomes clear that quantum geometry, as an open system, does not exhibit unitary evolution, and its Hilbert space is incomplete.

To address this issue, we employed Krein space quantization to construct a complete quantum space \cite{gareta00,ta22} and utilized the ambient space formalism to establish an observer-independent formulation of quantum geometry \cite{taqg}. This approach allows for the fundamental definition of the quantum state of the universe, which encompasses both geometrical fields and matter-radiation fields, ensuring that its evolution is unitary \cite{taqg}.

Since quantum gravity is an open system that interacts with the matter-radiation fields embedded within it, this approach can be extended to open quantum systems in quantum computation. We aim to apply Krein space method to define new quantum algorithms for use in quantum computation. For a recent review of existing quantum algorithms, see \cite{port}.

In classical gauge-invariant field theories, the Green's function, which can be interpreted as the inverse of the field equation \eqref{grfu}, cannot be determined due to the presence of zero modes. To address this problem in theoretical physics at the classical level, it is necessary to fix the gauge \cite{bailse}. This issue mirrors challenges in computational algorithms involving matrices with zero determinants, as well as in non-unitary evolution systems or open quantum systems. At the quantum level, achieving covariant quantization requires employing indefinite metric quantization \cite{stroch} or Krein space quantization \cite{taqg,taga22}. Gauge fixing as a constraint has already been used in computer science (see section \ref{constraint}); here, we aim to use the Krein method for the first time to address such issues in the context of quantum computation (see section \ref{similarity}). 

In quantum computations \cite{Nielsen}, qubits or qudits, representing quantum states, interact with the environment, rendering them quantum open systems. However, the time evolution of a quantum open system is inherently non-unitary, which poses challenges for quantum computation, where unitary transformations are fundamental \cite{shsnm}.  Furthermore, in certain cases where the determinant of a matrix is zero, the system cannot be adequately described using a unitary transformation or unitary gate. To address this issue, three approaches are commonly employed: Quantum Singular Value Decomposition (QSVD),  Sz.-Nagy Dilation, and  Unitary Operator Decomposition. For a comprehensive review of these methods, see \cite{mazzioti2}. 

There are two prevalent perspectives on quantum theory \cite{taqg0}. The first is the classical perspective, where the physical system is described using spinor or tensor fields, as in effective field theory, and the corresponding quantum algorithms can be executed on classical computers. The second is the quantum perspective, in which the system is characterized by its quantum state, and the corresponding quantum algorithms can be executed only on quantum computing hardware.

Effective field theory incorporates quantum corrections into classical field theory, enabling quantum computations to be simulated on classical computers. This approach may improve the precision of classical calculations by integrating quantum insights. While sophisticated, it serves as a practical bridge until scalable quantum computing hardware becomes available. However, an important question arises: do quantum algorithms based on effective field theory outperform classical algorithms when implemented on classical computّers? Addressing this question lies beyond the scope of this article and will be explored in future research.

In Section \ref{Secmaen}, we review key results from quantum field theory that are fundamental to our discussion. This includes the principle of least action, which serves as a central theme, along with the concepts of free fields, interaction fields (gauge theory), and quantum states. In Section \ref{qccha}, we address challenges in quantum computation and highlight their similarities to problems encountered in theoretical physics. Constrained physical systems are examined in Section \ref{constraint}. Section \ref{similarity} introduces a new algorithm formulated within Krein space. Finally, concluding remarks are provided in Section \ref{conclu}. In Appendix \ref{appendix}, we present a simple example to enhance the clarity and comprehensibility of the paper.

%%%%%%%%%%%%%%%%%%%%%%%%%%%%%%%%%%%%%%
\section{Important Notions} \label{Secmaen}

Classical physical systems are modeled using particles and tensor fields. In the framework of first quantization, a particle is described by tensor or spinor fields, collectively forming the basis of classical field theory \cite{taja,takook2}. In classical field theory, including first quantization, the field equation is derived from the principle of least action and the Lagrangian density. The principle of least action is expressed as:
$$ \label{sode}    S= \int \ud\mu(x) \mathcal{L}_c(\Phi, \partial \Phi; m,e),\;\;  \delta S=0\; ,$$
which leads to the field equation:
\b \label{cfe} D(x;m) \Phi(x)= f(\Phi,e)\,,\e
where $x^\mu$ represents the spacetime point, $\Phi(x)$ denotes the tensor or spinor field, $D$ is a second-order differential operator, and $\ud\mu(x)$ is the invariant volume element of spacetime. The term $f(\Phi,e)$ contains nonlinear contributions, explicitly determined by the interaction potential and gauge theory. The classical Lagrangian density, $ \mathcal{L}_c(\Phi, \partial \Phi; m,e)$, depends on the field $\Phi$, its derivatives $\partial \Phi$, the mass parameter $m$ (representing the energy scale of the physical system), and the coupling constant $e$ (defining the strength of the interaction). The kinematic terms in the Lagrangian density are derived from the symmetry group of spacetime, such as the Poincar\'e group or de Sitter group. In contrast, the dynamic (interaction) terms are obtained from gauge theory or local symmetry principles \cite{wei,ta14}.

The solution of the homogenous equation, $D(x;m)\Phi_0(x)=0$, allows us to calculate the Green's function for the corresponding equation:
\b  \label{grfu}  D(x;m)G(x,y)= \delta(x,y),\e
where $\delta(x,y)$ is the Dirac delta function with the following properties:
$$ \int \ud\mu(y) \delta(x,y)=1,\;  \int \ud\mu(y) \delta(x,y)f(y)=f(x)\, .$$
 In general, $G$ can be considered the inverse of  $D$ ($G \equiv D^{-1})$. Consequently, the general solution of the field equation \eqref{cfe} can be expressed as a perturbation series:
 $$ \Phi(x)=\Phi_0(x)+ \int \ud\mu(y) G(x,y)  f(\Phi(y),e)=\Phi_0(x)+ \int \ud\mu(y) G(x,y)  f(\Phi_0(y),e)+\cdots\,. $$
It is important to note that it is not always possible to invert the operator $D(x;m)$ and obtain the invariant two-point function $G(x,y)$ in classical field theory, due to zero eigenvalues, as in the case of massless fields in gauge theory, $D(x;m=0)$ \cite{bailse}, zero-mode problems in linear gravity, massless minimally coupled scalar fields in de Sitter space \cite{ilto}.

In classical gauge theory, this problem is addressed by fixing the gauge, which involves adding a constraint or term to the Lagrangian density:
 \b \label{gft}  \mathcal{L}_c(\Phi, \partial \Phi; m, e, \lambda)=   \mathcal{L}_c(\Phi, \partial \Phi; m, e)+\lambda C(\Phi),\e
 where $C(\Phi)$ is the constraint equation and $\lambda$ is the gauge-fixing parameter. The principle of least action then leads to a modified field equation, allowing one to obtain the inverse of $D(x; m=0, \lambda)$ in this context. However, in quantum field theory, constructing a covariant quantization in such cases often leads to symmetry breaking. These issues arise across various domains of theoretical physics, such as quantum electrodynamics, quantum Yang-Mills theory, massless minimally coupled scalar fields in de Sitter space, and quantum gravity. To resolve these problems, indefinite metric quantization \cite{stroch} and/or Krein space quantization must be employed \cite{taqg, gareta00}.

There are two dominant perspectives on quantum field theory, referred to as the classical view and the quantum viewpoint \cite{taqg0}. In the classical view, the physical system is modeled using tensor or spinor fields, and this perspective can be formulated within the framework of effective field theory. In effective field theory, quantum corrections are incorporated into the classical Lagrangian density as follows:
\b \label{qld} \mathcal{L}_q(\Phi, \partial \Phi; m, e, \lambda,\hbar)= \mathcal{L}_c(\Phi, \partial \Phi; m, e,\lambda)+ \hbar \mathcal{L}^{(1)}+\hbar^2 \mathcal{L}^{(2)}+\cdots\, ,\e
where $\mathcal{L}_c$ represents the classical Lagrangian density, $\hbar \mathcal{L}^{(1)}$ is the first-order quantum correction, and higher-order corrections $(\hbar^2 \mathcal{L}^{(2)}$, etc.) follow in the series. The quantum corrections introduce linear terms into the operator $D(x; m, \lambda, \hbar)$ and nonlinear terms into the second part of the field equation: 
\b \label{firper}  D(x; m, \lambda, \hbar)  \Phi(x)= g(\Phi,e,\hbar)\,, \e
where $D(x; m, \lambda, \hbar)$ incorporates the linear quantum effects, and $g(\Phi, e, \hbar)$ includes the nonlinear contributions arising from interactions and higher-order corrections.

In the second perspective, the physical system is represented by a quantum state $\vert \alpha \rangle$, which resides in the Hilbert space of the physical system, $\mathcal{H}$. The objective is to determine this quantum state and its evolution over time. By applying the principles of quantum field theory in Minkowski space and specifying the initial conditions, $|\alpha, \text{in}\rangle$, the quantum state's evolution can be predicted with certainty. This determinism is governed by the unitarity principle, which is fundamental in quantum computation:
\b \label{secper} \vert \alpha ,out \rangle=U(out,in; H) |\alpha ,in\rangle,\;\;\; UU^\dag=U^\dag U=\mbox{{ I}}\,,\e
where $\vert \alpha, \text{out} \rangle$ represents the final state, $U$ is the unitary evolution operator, and $\text{I}$ is the identity operator in the Hilbert space $\mathcal{H}$. The Hamiltonian operator $H$, which governs the system's dynamics, can be derived from the Lagrangian density $\eqref{qld}$. The Hamiltonian $H$ is a Hermitian operator, satisfying $H = H^\dagger$. In general, calculating the unitary evolution operator $U$ can be highly complex and may require expressing it as a series involving unitary irreducible representations of the Poincar\'e group or the de Sitter group \cite{wei,ta14}. However, for simpler physical systems, the unitary evolution operator can sometimes be determined explicitly.

The classical view of quantum theory focuses on understanding and predicting the behavior of tensorial or spinorial fields, while the quantum perspective aims to determine the quantum state and its evolution. These two perspectives give rise to two distinct classes of quantum algorithms. The first perspective, described by equation \eqref{firper}, leads to algorithms that can be programmed in bit language and executed on classical computers. In contrast, the second perspective, described by equation \eqref{secper}, necessitates programming in qubit or qudit language and requires quantum hardware to execute the corresponding quantum algorithm.

It is noteworthy that the concepts of gauge transformation in ambient space formalism \cite{taga22} and Krein space quantization \cite{gareta00} enable the construction of quantum geometry \cite{taqg} and the de Sitter supergravity \cite{taksup}. Similarly, in quantum computing, the principles of quantum singular value transformation \cite{svt} and higher-dimensional space \cite{martyn} have the potential to unify all quantum algorithms within the framework of the second perspective on quantum field theory. These ideas show a remarkable similarity to ambient space formalism and Krein space quantization. Consequently, ambient space formalism and Krein space quantization could serve as powerful tools for achieving a unification of quantum algorithms.

In quantum field theory, unitarity is a fundamental property of the S-matrix (scattering matrix), which ensures that the total probability of all possible outcomes of a quantum process is conserved. This principle, known as probability conservation, is inherently linked to energy conservation. If unitarity were violated in a theory, such as quantum gravity, it would imply that the physical system is not closed and that energy is not conserved, indicating the presence of a quantum open system. In such systems, interactions occur with external entities or fields. For instance, in quantum gravity, the gravitational field interacts with matter and radiation fields, making it a quintessential example of a quantum open system.

In quantum information theory, unitarity guarantees that no information is lost during the evolution of a quantum system. A violation of unitarity in this context indicates that the chosen Hilbert space is incomplete for the system under consideration. This implies that information is leaking outside the Hilbert space, necessitating its augmentation with additional basis elements. For more details on a related problem, see \cite{tagahu}.

%%%%%%%%%%%%%%%%%%%%%%%%%%%%%%%%%%%%%%%

\section{Computational challenges} \label{qccha}

In computer science, many problems can be reformulated as finding the solution to the following matrix equation:
\b \label{mle}  Av=b(v)\,  ,\e
where $A$ is a matrix, and $v$ and $b$ are vectors. In general, $b(v)$ may depend on $v$. This equation is analogous to equations \eqref{cfe} and \eqref{firper}, and it can be extended to the form of equation \eqref{secper}. Here, $b(v)$ corresponds to the source term $f(\Phi, e)$, $A$ is analogous to the operator $D$, and the vector $v$ represents the tensor or spinor field $\Phi$. In the generalized case, $v$ corresponds to the quantum state $\vert \alpha \rangle$.

If we assume that $b$ is independent of $v$ (i.e., a linear equation) and $\det A \neq 0$, then the solution can be straightforwardly obtained using the inverse of $A$ \cite{hhlqa}:
$$ v=A^{-1}b.$$
In computer science, many problems in numerical analysis, scientific computing, and data science reduce to solving linear equations. However, complications arise when some eigenvalues of $A$ are zero, rendering $A$ non-invertible (i.e., $\det A = 0$). By diagonalizing $A$ as $Au_i = \lambda_i u_i$ and normalizing and rearranging the eigenvalues, $A $ can be formally expressed as:
\b \label{sigm}
A= \left(
\begin{array}{ccc}
( \lambda_i)_{m\times m} & 0 _{m\times n}  \\
0_{n\times m}  &   0_{n\times n}
\end{array}
\right)\,,    \;\; \lambda_i  \neq 0, \;\; i=1, \cdots, m,
\e
where $(\lambda_i)_{m \times m}$ is an $m \times m$ diagonal matrix containing the non-zero eigenvalues. If there are $n$ zero eigenvalues, the matrix $A$ exhibits symmetry under $SU(n)$, corresponding to these zero-mode eigenvalues. This scenario frequently arises in applications such as signal and image processing, as well as machine learning problems. It is analogous to the gauge-invariant issues discussed in the preceding section.

To compute the eigenvalues of a matrix $A$, the singular value decomposition (SVD) is commonly used. SVD is a factorization of a real or complex matrix into three components: a rotation, followed by a rescaling, and then another rotation:
\b \label{svd} A= U\Sigma V^\top \,,\e
where $U$ and $V$ are orthogonal matrices ($UU^\top = I = VV^\top$), and $\Sigma$ is a diagonal matrix containing the singular values, similar in form to the matrix \eqref{sigm}. However, due to the presence of zero-mode eigenvalues, the inverse of the matrix $\Sigma$ does not exist, and consequently, the matrix $A$ is not unitary.

In this case, to obtain the solution of the linear equation \eqref{mle}, the Moore-Penrose inverse $A^+$, often called the pseudoinverse, is commonly used. It satisfies the following four criteria:
$$ AA^+A=A ,\;\;  A^+AA^+=A^+,\;\;\; (A^+A)^\dag=A^+A  ,\;\; (AA^+)^\dag=AA^+. $$
Using the SVD of $A$ from \eqref{svd}, the pseudoinverse $A^+$ can be expressed as $A^+ = V\Sigma^+ U^\top$, where $\Sigma^+$ is obtained by replacing the nonzero singular values in $\Sigma$ with their multiplicative inverses. If the linear system has any solutions, they are given by:
$$  v=A^{+}b+\left[I-A^{+}A\right]w\, , $$
where $w$ is an arbitrary vector. Existence, uniqueness, and stability are three conditions for a well-posed problem, which can be used to determine the possible choices for $w$. Solutions exist if and only if $AA^+b = b$. However, this condition is not always satisfied in real-world problems, such as those encountered in signal and image processing or machine learning. The arbitrary vector $w$ reflects the non-uniqueness inherent in the solution, analogous to the arbitrariness issue in gauge theory. This issue must be resolved by imposing auxiliary constraints, which will be discussed in the next section (see Section \ref{constraint}).

Another problem arises because $A$ is not unitary and therefore cannot be directly implemented on quantum processors. To address this issue, methods such as Quantum Singular Value Decomposition, Sz.-Nagy dilation, or Unitary Operator Decomposition are often employed. For a recent review, see \cite{mazzioti2}.

In general, these methods construct a dilated unitary matrix for quantum computation by embedding  $A$  into a larger matrix  $U_A $ that extends  $A $ into a unitary form. This involves constructing a unitary matrix  $U_A $ in the form:
\b \label{dilated}
U_A = \begin{pmatrix}
A & \sqrt{I - AA^\dag} \\
\sqrt{I - A^\dag A} & -A^\dag
\end{pmatrix},
\e
where \( \sqrt{I - AA^\dag} \) is known as the defect operator. The matrix  $A$  and the unitary  $U_A$  satisfy:
\b \label{totals} A \in \mathcal{B}(\mathcal{H}), \quad U_A \in \mathcal{B}(\mathcal{H}_T),  \quad  \mathcal{H}_T = \mathcal{H} \oplus \mathcal{H}\, , \e
where $\mathcal{H}_T $ denotes the total Hilbert space and  $\mathcal{B}(\mathcal{H}) $ denotes the bounded linear operators on $ \mathcal{H}$. Here,  $I$  is the identity operator in $\mathcal{B}(\mathcal{H}) $.  The matrix  $U_A$  is unitary because it satisfies \( U_A U_A^\dag = U_A^\dag U_A = \mathbb{I} \), where  $\mathbb{I}$  is the identity operator in  $\mathcal{B}(\mathcal{H}_T)$.

Using this approach, the diagonal operator $\Sigma$ can also be dilated into a unitary operator:
$$ 
U_\Sigma= \left(
\begin{array}{ccc}
\Sigma &  \sqrt{1-\Sigma \Sigma^\dag}   \\
\sqrt{1-\Sigma \Sigma ^\dag}  &   -\Sigma^\dag
\end{array}
\right)\, . $$
This formulation leverages the unitary and invertible nature of $U_\Sigma$ to circumvent the issues arising from $\Sigma$ being non-invertible.

Some authors prefer to work with the diagonalized form, which can be expressed as follows \cite{svt, ohetal}:
\b \label{svegva}
U_\Sigma^d= \left(
\begin{array}{ccc}
\Sigma_+  &  0   \\
0 &   \Sigma_-
\end{array}
\right)\, ,
\e
where the singular values of the unitary diagonal matrix are given by $\Sigma_{i,\pm} = \sigma_i \pm i\sqrt{1 - \sigma_i^2}$. These eigenvalues lie on the unit circle in the complex plane, as required for a unitary matrix. This approach bears resemblance to the $i\epsilon$-prescription in quantum field theory, which avoids singularities in the Green's function \cite{bailse}.

In the Unitary Operator Decomposition method, the matrix $A$ is first decomposed into its Hermitian and anti-Hermitian components:
$$A=A_h+A_a,$$
where both $A_h$ and $A_a$ operate on a Hilbert space of dimension $N$. A unitary matrix $U(A_h + A_a)$ is then constructed from these components, acting on a quantum state $\vert \alpha \rangle$ that resides in an enlarged space with four times the original Hilbert space dimension, $4N$, i.e., $\vert \alpha \rangle \in \mathcal{H} \oplus \mathcal{H} \oplus \mathcal{H} \oplus \mathcal{H}$. For further details, see \cite{shsnm2}. The enlargement of the space in this method is analogous to Krein space quantization, which will be discussed in Section \ref{similarity}.

%%%%%%%%%%%%%%%%%%%%%%%%%%%%%%%%%%%%%%
\section{Constraints}\label{constraint}

Due to the presence of zero-mode solutions in the equation \eqref{mle}, it is impossible to obtain a unique solution without imposing constraints to identify the optimal or physically meaningful solution. This issue closely resembles the Gribov Ambiguity in gauge theory \cite{bailse}.

Gribov ambiguity arises when attempting to isolate the physical gauge degrees of freedom in non-Abelian gauge theory by imposing a covariant gauge constraint. Unfortunately, the solutions to the gauge constraint are not unique, leaving redundant gauge degrees of freedom, known as Gribov copies, unfixed. A traditional approach to partially address the Gribov problem is to restrict the space of gauge orbits to a bounded region called the Gribov region \cite{bailse}.
 
In gauge theory, constraints are used to determine the Green's function in classical field theory and to separate the physical degrees of freedom from the non-physical ones in quantum field theory \cite{bailse}. Similarly, in computer science, constraints are applied to first compute the inverse of a matrix and subsequently to identify a unique and optimal solution that best aligns with the observations.

For instance, the constrained least-squares problem is an optimization problem aimed at finding the best-fit solution to a set of linear equations \eqref{mle} where the matrix A is singular, subject to specific constraints on the solution. This ensures that the solution is both mathematically and practically meaningful.

Given a matrix $A \in \mathbb{R}^{m \times n}$ and a vector $b \in \mathbb{R}^{m}$, the objective is to find a vector $v \in \mathbb{R}^{n}$ that minimizes the Euclidean norm of the residual vector $(Av - b)$, {\it i.e.}, $ \min_{v} \|Av - b\|^2 $, while satisfying certain constraints on $v$.
The constraints can take various forms, such as equality constraints ($ Cv = d $ where $C \in \mathbb{R}^{p \times n}$ and $d \in \mathbb{R}^{p}$), inequality constraints ($ Gv \leq h $ where $G \in \mathbb{R}^{q \times n}$ and $h \in \mathbb{R}^{q}$), or bounds on the variables ( $ l \leq v \leq u $ where $l$ and $u$ are vectors representing the lower and upper bounds on $ v$). 

For example, the minimization of the residual $(Av - b)$ subject to an exact constraint $(Cv = d)$ can be expressed in the Lagrangian formalism as:
$$  \mathcal{L}(v, \lambda) = \|Av - b\|^2 + \lambda (Cv - d) ,$$
where $\lambda$ are the Lagrange multipliers, which play a role analogous to the gauge-fixing parameter in gauge theory \eqref{gft}.
 
To find the optimal $v$ and $\lambda$, take the partial derivatives of $\mathcal{L}$ with respect to $v$ and $\lambda$ and set them to zero:
$$ \frac{\partial \mathcal{L}}{\partial v} = 2A^T(Av - b) + C^T \lambda = 0, \;\;  \frac{\partial \mathcal{L}}{\partial \lambda} = Cv - d = 0. $$
Solve these equations simultaneously to obtain the solution. The constrained least-squares problem is a cornerstone of applied mathematics and optimization, with a wide array of practical applications in fields such as machine learning, engineering, economics, and data science. There are the alternative methods trying to optimize the solution such as:
$$L=\|Av-b\|^2+\lambda \|Cv-d\|^2$$
or also other norms, fro review see \cite{amdjafari}.

%%%%%%%%%%%%%%%%%%%%%%%%%%%%%%%%%%%%%%

\section{Krein space method} \label{similarity}

In theoretical physics, indefinite metric quantization arises in the context of quantizing gauge theories and gauge gravitation. In such cases, the space of physical states is incomplete under the action of the algebra of field operators. To achieve completeness, the state space must be extended to include states with negative norms. One example of such an extended space, which we discuss here, is the Krein space.

A Krein space is a complex linear space with an indefinite inner product, denoted by  $\langle \cdot, \cdot \rangle $. It can be decomposed into an orthogonal direct sum of two subspaces: one with a positive-definite inner product and the other with a negative-definite inner product \cite{gareta00}:
\b \label{onepks} \mathcal{K} \equiv \mathcal{H} \oplus \mathcal{H}^*\equiv K_+ \oplus K_-\,,\e
where $\mathcal{H} (K_+)$ is a Hilbert space, and $\mathcal{H}^* (K_-)$ denotes its dual space. This symmetry is referred to as the fundamental decomposition. 

Krein space is similar to the total Hilbert space $\mathcal{H}_T = \mathcal{H} \oplus \mathcal{H}$, as defined in \eqref{totals}. Using the decomposition \eqref{onepks} of the Krein space $\mathcal{K}$, a Hilbert inner product $(\cdot, \cdot)$ can be defined as:
$$  (a, b)= \langle a_+, b_+ \rangle - \langle a_-, b_- \rangle,$$
where $a = a_+ + a_- $ and $b = b_+ + b_-$. The orthogonal projections onto $K_+$ and $K_-$ are denoted by $P_+$ and $P_-$, respectively. For the operator $J = P_+ - P_-$, called the fundamental symmetry, the following relations hold:
$$\langle a, b \rangle=  (Ja, b)=(a, Jb),$$
where $J^2 = I$ and \(J = J^\dag\). The operator $J$, also known as the fundamental symmetry or signature matrix, distinguishes between the components in $\mathcal{H}$ and $\mathcal{H}^*$. It is a diagonal operator with entries $\pm 1$, reflecting the positive and negative norm components of the space.  Conversely, given a total Hilbert space $\mathcal{H}_T \equiv \mathcal{H} \oplus \mathcal{H}$ and an operator $J$ with the aforementioned properties, an indefinite inner product space $\mathcal{K}$ can be constructed. Thus, one can transition between the two formalisms.

In Krein space, the concept of unitarity is generalized to what is known as a Krein-unitary operator. Krein-unitary operators satisfy conditions that ensure the preservation of the indefinite inner product defined by the metric operator  $J$ . An operator  $K$  in this space is Krein-unitary if it preserves the indefinite inner product, as specified by the condition:
\b \label{ku}   K^\dagger J K= J \,. \e

Similarly, an operator  $H$  in this space is considered Krein-Hermitian if it satisfies the condition:
\b \label{kh} H^\dag =JHJ\,, \e
which ensures that the structure of the Krein space is preserved under its action. These definitions generalize the conventional notions of unitarity and Hermiticity to the context of Krein spaces.

The Krein space can be utilized to address problems involving matrices with zero determinants or, more specifically, to study open quantum systems. It is important to note that the extension from a Hilbert space to a Krein space is not unique, and multiple possibilities exist. In this context, we consider a simple toy model. For simplicity, and in applications related to matrix inversion, we also focus on a finite-dimensional Krein space, where operators are represented by matrices  $A_k$  and states by vectors  $\vert n_k \rangle $.

For a matrix $A$ acting on the Hilbert space $\mathcal{H}$, we propose the following ansatz for its dilation into the Krein space $\mathcal{K}$ in the form:
$$ A_k= \begin{pmatrix}
   A & {\bf \cdot} \\
   {\bf \cdot} & {\bf \cdot}
   \end{pmatrix} , \;\; A_k\vert n_k \rangle=\vert n'_k \rangle\in \mathcal{K} \equiv \mathcal{H} \oplus \mathcal{H}^*,$$
where the symbol "$\mathbf{\cdot}$" denotes arbitrary elements of the matrix.

If we require the matrix  $A_k$  to be unitary or Hermitian, it must satisfy the conditions defined by equations $\eqref{ku}$ or $\eqref{kh}$, respectively, thereby imposing constraints on the arbitrary elements.  Additionally, we impose the condition  $\det A_k \neq 0$  to define these elements, even in cases where  $\det A = 0 $.

Inspired by the fundamental symmetry of Krein space and the condition  $\det A_k \neq 0$, the matrix  $A_k$  may be chosen in the following simple form:
\b  \label{krmat}  A_k= \begin{pmatrix}
   A & \mu \mbox{I} \\
  \mu \mbox{I} & A
   \end{pmatrix} , \e
where  $\mathrm{I}$  is the identity operator in the Hilbert space  $\mathcal{H}$. The parameter  $\mu$  is a free variable, referred to as the regularization parameter, which can be adjusted to ensure  $\det A_k \neq 0$.

It is important to note that for an ill-posed problem in Hilbert space, where  $\det A = 0$, it is always possible to choose  $\mu$  such that  $\det A_k \neq 0 $:
$$ \det {\begin{pmatrix}A&\mu \mbox{I} \\\mu \mbox{I} &A\end{pmatrix}}=\det(A-\mu \mbox{I} )\det(A+\mu \mbox{I} )\neq 0.$$
Determining the regularization parameter is typically the most crucial and time-consuming step in the standard approach \cite{shxi}. However, in our model, this process is significantly simplified and can be performed using the variational method.

The matrix structure \eqref{krmat} is not unique and represents the interactions between the  $K_+$  and  $K_-$  subspaces within a Krein space. In this context, the parameter  $\mu$  governs the strength of the interaction between  $K_+$  and  $K_- $. For any non-zero  $\mu$ , it can be demonstrated that these interactions can render the full matrix invertible, even when the matrix in the  $K_+$  subspace is singular. When  $\mu = 0$, there is no interaction, and the matrix becomes block diagonal. As  $|\mu |$  increases, the strength of the interaction correspondingly increases.

The matrix \eqref{krmat} exhibits a symmetric structure, in which the subspaces $K_+$ and $K_-$ possess identical internal configurations and symmetric interactions. This structure may represent a system composed of two identical subsystems, coupled through a single parameter $\mu$. The coupling preserves the internal integrity of each subsystem while enabling their interaction. Since $\det A_k \neq 0$, the inverse $A_k^{-1}$ exists and can be used to solve the extended system $A_k \mathbf{v}_k = \mathbf{b}_k$, yielding $\mathbf{v}_k = A_k^{-1} \mathbf{b}_k$ (for a simple example, see Appendix~\ref{appendix}). The parameter $\mu$ can be optimized to achieve the best fit with observational data; this will be discussed in detail in a forthcoming paper~\cite{taja2}.

%%%%%%%%%%%%%%%%%%%%%%%%%%%%%%%%%%%%%%
\section{Conclusion}  \label{conclu}

The de Sitter universe plays an important role in both the early and present-day descriptions of the cosmos. Quantum field theory in this spacetime provides a framework for addressing unresolved problems in theoretical physics, such as quantum geometry and quantum Yang-Mills theory. Recently, quantum field theory in de Sitter space has also been explored in the study of neural networks within the context of quantum computational methods~\cite{lly}.

This study highlights the intricate relationship between concepts in theoretical physics-such as gauge theory and Krein space quantization-and principles in quantum computing, including singular value decomposition, unitary operator decomposition, and dilated unitary matrices. By drawing parallels between challenges encountered in theoretical physics and those arising in computer science, we demonstrate how techniques from the former can be leveraged to address computational problems. For instance, ill-conditioned linear systems of the form $A v = b$, where $\det(A) = 0$, can be effectively resolved through dilation into a Krein space. This approach not only regularizes the problem but also provides a promising path toward developing new quantum algorithms grounded in physical principles.

\vspace{1.0cm}
{\bf{Acknowledgments}}: The author wishes to express particular thanks to A.M. Djafari and A. Djannati-Atai and S. Rouhani for their discussions. The author would like to thank Coll\`ege de France, Universit\'e Paris Cit\'e, and Laboratorie APC for their hospitality.

%%%%%%%%%%%%%%%%%%%%%%%%%%%%%%%%%%%%%%%

\begin{appendix}

\section{Solving a Singular Linear System via Krein Regularization} \label{appendix}
\addcontentsline{toc}{section}{Appendix A: Worked Example of Krein Regularization}

To demonstrate the practical application of the proposed Krein space algorithm, we consider a simple toy example involving a singular linear system of equations in two-dimensional space ($1$-qubit), which can be readily generalized to an $N$-dimensional space (corresponding to d-qubits, where $N = 2^d$).

Let us take:
\begin{equation}
A \mathbf{v} = \mathbf{b}, \quad \text{where} \quad 
A = \begin{pmatrix} 1 & 1 \\ 2 & 2 \end{pmatrix}, \quad 
\mathbf{b} = \begin{pmatrix} 2 \\ 4 \end{pmatrix}.
\end{equation}
This system is ill-conditioned, with $\det(A) = 0$, meaning the inverse of $A$ does not exist and the solution is not unique.

To regularize this system, we embed it into a Krein space by constructing the following extended matrix:
\begin{equation}
A_k = \begin{pmatrix} A & \mu I \\ \mu I & A \end{pmatrix}, 
\end{equation}
where $I$ is the $2 \times 2$ identity matrix and $\mu$ is a nonzero regularization parameter to be determined. Explicitly, this gives:
\begin{equation}
A_k = \begin{pmatrix}
1 & 1 & \mu & 0 \\
2 & 2 & 0 & \mu \\
\mu & 0 & 1 & 1 \\
0 & \mu & 2 & 2
\end{pmatrix}.
\end{equation}

The extended system becomes:
\begin{equation}
A_k \mathbf{v}_k = \mathbf{b}_k, \quad \text{where} \quad 
\mathbf{b}_k = \begin{pmatrix} \mathbf{b} \\ \mathbf{0} \end{pmatrix} 
= \begin{pmatrix} 2 \\ 4 \\ 0 \\ 0 \end{pmatrix}.
\end{equation}
However, the solution vector $\mathbf{v}_k \in \mathcal{K} $ depends on the regularization parameter $\mu$, {\it i.e.}, $\mathbf{v}_k(\mu)$. This parameter must be determined using an appropriate regularization method in order to extract physically meaningful results.

As long as $\mu \ne 0$, the determinant of $A_k$ satisfies:
\begin{equation}
\det(A_k) = \det(A - \mu I)\cdot \det(A + \mu I) \neq 0,
\end{equation}
ensuring that $A_k$ is invertible. We can thus solve for:
\begin{equation}
\mathbf{v}_k = A_k^{-1} \mathbf{b}_k,
\end{equation}
and extract the physical solution by projecting onto the original Hilbert space:
\begin{equation}
\mathbf{v} = P_+ \mathbf{v}_k,
\end{equation}
where $P_+$ is the projection operator onto the  positive-definite subspace $K_+ = \mathcal{H}$ subspace. The parameter $\mu$ acts as a tunable regularization parameter. Its optimal value can be determined empirically, or through variational techniques, depending on the physical or computational context.

This example clearly illustrates how Krein space dilation regularizes a singular system and enables meaningful quantum algorithm construction even when standard methods fail. The coupling parameter $\mu$ governs the interaction between $K_+$ and $K_-$, allowing fine-tuning for optimal stability or empirical alignment.

\end{appendix}

%%%%%%%%%%%%%%%%%%%%%%%%%%%%%%%%%%%%%%%%

\end{document}